\documentstyle[eqsecnum,aps,preprint,tabls,psfig]{revtex}

\newcommand{\be}{\begin{eqnarray}}
\newcommand{\ee}{\end{eqnarray}}
\begin{document}
\tighten
\title{Mesons in (2+1) Dimensional Light Front QCD. II. Similarity Renormalization
Approach}
\author{
{Dipankar Chakrabarti}\thanks{e-mail: dipankar@theory.saha.ernet.in} and
{A. Harindranath}\thanks{e-mail: hari@theory.saha.ernet.in} \\
{\it Saha Institute of Nuclear Physics, 1/AF Bidhan Nagar, 
	Calcutta 700064 India}}
\date{August 8, 2001}
\maketitle
\begin{abstract}

Recently we have studied the Bloch effective Hamiltonian approach to bound 
states in 2+1 dimensional gauge theories. Numerical calculations were 
carried out to investigate the vanishing energy denominator problem. In this 
work we study similarity renormalization approach to the same problem. By
performing analytical calculations with a step function form for the
similarity factor, we show that in addition to curing the vanishing energy 
denominator problem, similarity approach generates linear confining 
interaction for large transverse separations. However, for large longitudinal 
separations, the generated interaction grows only as the square root of the 
longitudinal separation and hence produces violations of rotational symmetry 
in the spectrum. We  carry out numerical studies in the G{\l}azek-Wilson and 
Wegner formalisms and present low lying eigenvalues and wavefunctions. 
We investigate the sensitivity of the spectra to various parameterizations of 
the similarity factor and other parameters of the effective Hamiltonian,
especially the scale $\sigma$. Our results illustrate the need for higher 
order calculations of the effective Hamiltonian in the similarity 
renormalization scheme.
\end{abstract}  
\vskip .1in
\noindent{PACS Numbers: 11.10.Ef, 11.10.Kk, 11.10.St, 11.15.Tk, 12.38.Lg} 
\vskip .2in
\vskip .2in
\section{Introduction}
Any attempt to solve bound state problems in quantum field theory using Fock
space based Hamiltonian methods immediately encounters a two-fold infinity
problem. In a relativistic system there are infinitely many energy scales and
there are an infinite number of particles. Typical Hamiltonians of interest
couple low energy scales with high energy scales which results in
ultraviolet divergences. Furthermore, Hamiltonian couples every particle
number sector allowed by symmetries and at strong coupling brute force 
particle truncation can fail miserably. 

Light front Hamiltonian poses special problems for the treatment of
ultraviolet divergences since the counterterms for transverse divergences 
in this case are in general expected to be nonlocal\cite{Wilson:1994fk} in the
longitudinal direction. In spite of the complexities due to renormalization,
light front approach to QCD is appealing from several physical aspects.
First and foremost is the fact that in a {\em cutoff theory} where one imposes a
cutoff on the minimum longitudinal momentum of the constituents, Fock vacuum
is the exact eigenstate of the Hamiltonian and there is hope that one may
find a direct link between QCD on one hand and constituent quark model and
quark parton model on the other hand.      

In the best of all possible worlds, one can attempt to diagonalize the
Hamiltonian in a single step. This is the spirit behind straightforward
implementations of the Discretized Light Cone Quantization (DLCQ) method 
(for a review see Ref. \cite{Brodsky:1998de}). DLCQ
has been quite successful in two dimensional models, but for QCD this
approach may
be quite ambitious. Recall efficient numerical
algorithms designed for matrix diagonalization where the matrix is first
brought to a tri-diagonal form which is then diagonalized\cite{Recipes}. In
the same spirit,     
to keep the complexities under control, a two step process to the Hamiltonian
diagonalization has been suggested\cite{Wilson:1994fk}. The proposal is based
on similarity renormalization scheme\cite{Similar} devised independently by
G{\l}azek and Wilson and Wegner. 

In the first step, by
integrating out the high energy degrees of freedom {\em perturbatively} one
arrives at an effective Hamiltonian which is in a band diagonal form. At
this step, one can identify the ultraviolet divergent part of the 
counterterms needed to be added to the
Hamiltonian to remove ultraviolet divergences. It is quite advantageous to
treat the ultraviolet divergences perturbatively especially in gauge
theories since one can avoid pitfalls of other effective Hamiltonian
approaches. For example, it is well-known that a simple truncation of the 
Fock space (like a Tamm-Dancoff truncation) leads to uncanceled divergences 
as a result of violations of gauge symmetry.       

In the second step, the effective Hamiltonian is diagonalized exactly. It is
important to note that the high energy states integrated out contain small
$x$ gluons, i.e., gluons
having small longitudinal momentum fraction. Since the vacuum is trivial, it
is hoped that, as a result of integrating out small $x$ gluons which are
sensitive to long distance physics on the light front, the effective 
Hamiltonian may contain
interactions responsible for low energy properties of QCD. Indeed
Perry\cite{Perry:1994kp} found a logarithmic confining interaction in the $q {\bar q}$
sector in the lowest order effective interaction. 

Initial bound state studies in the similarity renormalization approach
worked in either the non-relativistic limit\cite{Brisudova} or in the heavy
quark
effective theory formalism\cite{Zhang:1997dd} to investigate heavy-quark
systems. Only recently, work has
begun\cite{Allen:2000kx} in the context of glueball spectrum
to address many practical problems, especially the numerical ones that one faces 
in this approach. Since the conceptual and technical problems one encounters
in QCD are numerous, we have initiated a study of bound state problems in
QCD in 2+1 dimensions\cite{dh1}. Our main motivation is not the fitting of
data but a critical evaluation of the strengths and weaknesses of the
various assumptions and approximations made in the similarity approach.

In Ref. \cite{dh1} we studied  the meson sector of 2+1 dimensional 
light front QCD using a Bloch
effective Hamiltonian\cite{Perry:1994mn} in the first non-trivial order. The resulting two 
dimensional integral equation was converted into a matrix equation and solved 
numerically.  
The vanishing energy denominator problem which leads to severe 
infrared divergences in 2+1 dimensions was investigated in detail. 
We defined and studied 
numerically a reduced model which is relativistic, free from infrared 
divergences, and exhibits logarithmic confinement. The manifestation and  
violation of rotational symmetry as a function of the coupling was studied 
quantitatively. 
Our study 
indicated that in the context of Fock space based effective Hamiltonian
methods to tackle gauge theories in 2+1 dimensions, approaches like similarity
renormalization method may be mandatory  due to uncanceled infrared divergences 
caused by the vanishing energy denominator problem. 

In this work we study similarity renormalization approach 
in the first non-trivial order 
to the same problem. The plan of this paper
is as follows. In Sec. II we review the similarity renormalization approach
in the G{\l}azek-Wilson and Wegner formalisms. In Sec. III we present the
effective Hamiltonian in the $q {\bar q}$ sector. Three different
parameterizations of the similarity factor are discussed in Sec. IV. In Sec.
V we present analytical calculations with the step function similarity factor
which clearly shows the emergence of linear and square root confining
interactions for large transverse and longitudinal separations respectively. In
Sec. VI we perform numerical calculations and present low lying spectra and
wavefunctions. Sec. VII contains discussion and conclusions.  

\section{Similarity Renormalization Theory for the Effective Hamiltonian}
Realizing that a single step diagonalization of the Hamiltonian is too
difficult, one may aim for the next best thing, namely, bringing the
Hamiltonian to a band diagonal form. This is the method proposed by G{\l}azek
and Wilson and Wegner independently\cite{Similar}. 

Starting from a cutoff Hamiltonian $H_B$ which includes canonical terms and
counterterms we wish to arrive at an effective Hamiltonian $H_{\sigma}$
defined at the scale $\sigma$ via a similarity transformation 
\begin{eqnarray}
H_\sigma = S_\sigma ~ H_B ~ S_\sigma^\dagger
\end{eqnarray}
where $S_\sigma$ is chosen to be
unitary.

The boundary condition is $ Limit_{\sigma \rightarrow \infty} ~H_\sigma
= H_B$.

Introduce anti-Hermitian generator of infinitesimal changes of scale
$T_\sigma$ through
\begin{eqnarray}
S_\sigma = {\cal T}~ e^{\int_{\sigma}^{\infty}d\sigma' ~ T_{\sigma'}}
\end{eqnarray}
where $ {\cal T}$ puts operators in order of increasing scale. 

For infinitesimal change of scale, $  S_\sigma = 1 - T_\sigma ~ d\sigma$
and $S_\sigma^\dagger = 1+ T_\sigma ~ d\sigma$. Then we arrive at the
infinitesimal form of the transformation
\begin{eqnarray}
{d H_\sigma \over d\sigma} = [ H_\sigma, T_\sigma]. \label{diffform}
\end{eqnarray}
This equation which has been called the flow equation of the Hamiltonian is
the starting point of the investigations. 

The basic goal of the transformation $S_\sigma$ is that $H_\sigma$ should be
band diagonal relative to the scale $ \sigma$. Qualitatively this means that
matrix elements of $H_\sigma$ involving energy jumps much larger
than $ \sigma$ should be zero. $T_\sigma$ still remains arbitrary to a great
extent. It is instructive to go through the steps of the derivation which
leads to the G{\l}azek-Wilson choice.

We write $ H_B = H_{B0}+H_{BI}$ where $H_{B0}$ is the free part and $H_{BI}$
is the interaction part of the bare cutoff Hamiltonian.
A brute force way of achieving our goal is to {\it define} the matrix
elements $ H_{I \sigma ij} = f_{\sigma ij} H_{BI  ij}$ 
where we have introduced the function $ f_{\sigma ij} =f(x_{\sigma ij})$
with $x$ a function of  ${\sigma^2}$ and  $ \Delta M^2_{ij}$.     
The function $f(x)$ should be chosen
as follows:
\begin{eqnarray}
\nonumber
{\rm when} \ {\sigma^2 } >> \Delta M^2_{ij} ,
&&\quad  f(x) = 1\qquad \qquad \qquad
{\it (near\ diagonal\ region)}; \nonumber \\
{\rm when} \  {\sigma^2} << \Delta M^2_{ij}, 
&& \quad  f(x)
 =0  \qquad \qquad \qquad {\it (far \ off\  diagonal\  region)};
 \nonumber \\
{\rm in \ between}  \qquad \qquad  f(x) ~~~~&& ~~~~~
{\rm drops \ from \ 1 \ to \ 0}
\qquad \qquad  
{\it (transition \ region)}  .  
\end{eqnarray}
Here $\Delta M^2_{ij} (=M^2_i - M^2_j)$ denotes the difference of invariant masses of states
$i$ and $j$. Because of the properties of $f$, $H_{I \sigma ij}$ is band 
diagonal. What is wrong with such a choice of inserting form factors by hand
at the interaction vertices? First of all, we have simply discarded degrees
of freedom above $\sigma$. Secondly,
$H_\sigma$ will have very strong dependence on
$\sigma$. Thirdly, to ensure that $H_\sigma$ has no ultraviolet cutoff
dependence, $H_B$ should contain canonical and counterterms. But, in
light front Hamiltonian field theory, because of the complexities due to
renormalization, a priori we do not know the structure of counterterms.

Note that in the definition of $H_{\sigma}$ given in Eq. (\ref{diffform}) 
the form
of $T_{\sigma}$ is still unspecified. 
In fact, a wide variety of choices are
possible. In the following, we consider the choices made by G{\l}azek 
and Wilson and Wegner. The price we have to pay
for the use of flow equations is that it will generate complicated
interactions even if the starting Hamiltonian has only simple interactions.
For example, starting with a Hamiltonian which has only 2 particle
interaction, the transformation will generate  3 particle interactions, 4
particle interactions, etc.

\subsection{G{\l}azek-Wilson Formalism}

Writing $ H_\sigma = H_0 + H_{I\sigma}$, noting that the free Hamiltonian
$H_0$ does not depend on $\sigma$ and taking matrix elements in free
particle states, we have,
\begin{eqnarray}
[H_\sigma, T_\sigma]_{ij} = (P_i^- - P_j^-) T_{\sigma ij} +
[H_{I\sigma},T_\sigma]_{ij}
\end{eqnarray}
where $ H_0 \mid i \rangle = P_i^- \mid i \rangle$, etc. .
i.e.,
\be
{1 \over f_{\sigma ij}}{d H_{I \sigma ij} \over d\sigma} = 
{ 1 \over f_{\sigma ij}} [H_{I\sigma},T_\sigma]_{ij} + {1 \over f_{\sigma
ij}} (P_i^- - P_j^-) T_{\sigma ij}.
\ee
Since we want $H_{I \sigma ij}$ to be band diagonal, it is advantageous to
trade 
${1 \over f_{\sigma ij}}{d H_{I \sigma ij} \over d\sigma}$ for
${ d \over d \sigma} \left [ { 1 \over f_{\sigma ij}}H_{I\sigma ij}\right ]$
which on integration has the chance to ensure that $H_{I\sigma ij}$
is band diagonal, we use
\begin{eqnarray}
{ d \over d \sigma} \left [ { 1 \over f_{\sigma ij}}H_{I\sigma ij}\right ]
+ {1 \over f^2_{\sigma ij}}{df_{\sigma ij} \over d\sigma} H_{I\sigma ij} =
{ 1 \over f_{\sigma ij}}{dH_{I\sigma ij} \over d \sigma} 
\end{eqnarray}
and arrive at 
\begin{eqnarray}
{d \over d \sigma} \left [ { 1 \over f_{\sigma ij}} H_{I\sigma ij} \right ] &&=
 { 1 \over f_{\sigma ij}}(P_i^- -
P_j^-)T_{\sigma ij} \nonumber \\
&&~~ + { 1 \over f_{\sigma ij}} [H_{I\sigma},
T_{\sigma}]_{ij} - { 1 \over f^2_{\sigma ij}}{d f_{\sigma ij} \over d\sigma}
H_{I\sigma ij}.\label{me1}
\end{eqnarray}
Still $T_{\sigma ij}$ is not defined. We next convert this equation into two
equations, one defining the flow of $H_{I \sigma ij}$ 
and other defining $T_{\sigma ij}$.
Recalling the starting equation Eq. (\ref{diffform}) we add and subtract
$[H_{I\sigma},T_\sigma]_{ij}$ to the r.h.s. and arrive at  
\begin{eqnarray}
{d \over d \sigma} \left [ { 1 \over f_{\sigma ij}} H_{I\sigma ij} \right ] &&=
[H_{I\sigma},T_\sigma]_{ij} + { 1 \over f_{\sigma ij}}(P_i^- -
P_j^-)T_{\sigma~ij} \nonumber \\
&&~~ + { 1 \over f_{\sigma ij}}(1-f_{\sigma ij}) [H_{I\sigma},
T_{\sigma}]_{ij} - { 1 \over f^2_{\sigma ij}}{d f_{\sigma ij} \over d\sigma}
H_{I\sigma ij}.\label{me}
\end{eqnarray}
G{\l}azek and Wilson choose $T_\sigma$ to be
\begin{eqnarray}
T_{\sigma ij} = { 1 \over P_j^- - P_i^-}\left [ (1 - f_{\sigma ij})
[H_{I\sigma},T_\sigma]_{ij}- { d \over d\sigma} ({\rm ln}~f_{\sigma ij})
H_{I\sigma ij} \right ]. \label{gwt}
\end{eqnarray}
Then from Eq. (\ref{me}), we have,
\begin{eqnarray}
{dH_{I\sigma ij} \over d \sigma} = f_{\sigma ij}
[H_{I\sigma},T_{\sigma}]_{ij}. \label{gwh}
\end{eqnarray}
 
Integrating Eq. (\ref{gwh}) from $\sigma$ to $\infty$, we arrive at,
\begin{eqnarray}
H_{I\sigma ij} = f_{\sigma ij} \left [ H_{IB ij} - \int_\sigma^\infty
d\sigma' [H_{I\sigma'},T_{\sigma'}]_{ij} \right ]. \label{gwhs}
\end{eqnarray}

Note that $H_{I\sigma ij}$ is zero in the far off-diagonal
region. This is clear from the solution given in Eq. (\ref{gwhs}) since
$f(x)$ vanishes when $x \geq 2/3$. 

$T_{\sigma ij}$ vanishes in the near diagonal region. When $i$ is close to
$j$, $f_{\sigma ij}=1$ and both $(1 - f_{\sigma ij})$ and ${d \over
d\sigma}f_{\sigma ij}$ vanishes. It follows, then, from Eq. (\ref{gwt}) that
$T_{\sigma ij}$ vanishes in the near-diagonal region. This guarantees that a
perturbative solution to $H_{I\sigma ij}$ in terms of $H_{BI ij}$ will never
involve vanishing energy denominators.

Our next task is to derive the effective Hamiltonian to second order in
perturbation theory. 
Using
\begin{eqnarray}
H_{I\sigma ik}^{(1)}  \simeq f_{\sigma ik} H_{BI ik} 
\ee
and
\be
T_{\sigma kj}~ && \simeq {1 \over P_j^- - P_k^-} \left \{ - { d \over d \sigma} 
({\rm ln} ~f_{\sigma kj}) f_{\sigma kj} H_{BI kj} \right \}
\end{eqnarray}
in Eq. (\ref{gwhs}), a straightforward calculation leads to
\begin{eqnarray}
H_{I\sigma ij}^{(2)} = - \sum_k H_{BI ik} H_{BI kj} \left [
{g_{\sigma ijk} \over P_k^- - P_j^-} + { g_{\sigma jik} \over P_k^- - P_i^-}
 \right ], \label{gw2eh}
\end{eqnarray}
where
\begin{eqnarray}
g_{\sigma ijk} &&= f_{\sigma ij} ~ \int_{\sigma}^\infty d\sigma' ~
f_{\sigma' ik}~
{ d \over d \sigma'} f_{\sigma' jk}, \nonumber \\
g_{\sigma jik} &&= f_{\sigma ij}~ \int_{\sigma}^\infty d\sigma'~ f_{\sigma'jk}
{ d \over d \sigma'} f_{\sigma'ik}.\label{gfact}
\end{eqnarray}
We find that the effective Hamiltonian in similarity perturbation theory is a
modification of the effective 
Hamiltonian in Bloch perturbation theory\cite{Perry:1994mn}. 

\subsection{Wegner Formalism}
In the Wegner formalism\cite{Wegner:2000gi},
the flow equation is given by
\be
{d H(l) \over d l} = [ \tau (l),H(l)].
\ee 
Wegner chooses 
\be
\tau(l) = [H_d, H] = [H_d, H_r]
\ee
where $H_d$ is the diagonal part of the Hamiltonian and $H_r$ is the rest,
i.e., $H=H_d+ H_r$. Here the word diagonal is used in the particle number
conserving sense. It is important to note that $H_d$ is not the free part
of the Hamiltonian and both $H_d$ and $H_r$ depend on the length scale $l$.

The light front Hamiltonian has dimension of $(mass)^2$ and hence 
$\tau$ has the dimension of $(mass)^4$, $l$ has dimension of 
${ 1 \over (mass)^4}$.

Expanding in powers of the coupling constant, 
\be
H = H_d^{(0)}+ H_r^{(1)}+ H_d^{(2)} + H_r^{(2)} + \ldots
\ee
where the superscript denotes the order in the coupling constant,
\be
\tau(l) = [H_d^{(0)}, H_r^{(1)}] + [H_d^{(0)}, H_r^{(2)}] + \ldots ~ .
\ee  
Then, to second order,
\be
{d H \over d l} = [[H_d^{(0)}, H_r^{(1)}], H_d^{(0)}] + [[H_d^{(0)},
H_r^{(1)}], H_r^{(1)}] + [[H_d^{(0)}, H_r^{(2)}], H_d^{(0)}] + \ldots ~ .
\ee
Introduce\cite{Gubankova:1997ha} the eigenstates of $H_d^{(0)}$,
\be
H_d^{(0)} \mid i \rangle = P^{-}_{i} \mid i \rangle. 
\ee
Then, to second order,
\be
{d H_{lij} \over dl} = - (P^-_i - P^-_j)^2 H_{rij}^{(1)} + [ \tau_l^{(1)},
H_r^{(1)}]_{ij} - (P^-_i - P^-_j)^2 H_{rij}^{(2)} + \ldots~ .
\ee
To first order in the coupling,
\be
{ d H_{rij} \over dl} = - (P^-_i - P^-_j)^2  H_{rij}^{(1)}
\ee
which on integration yields
\be
H_{rij}^{(1)} (\sigma) = e^{ - {(P^-_i - P^-_j)^2 \over \sigma^4}}
H_{rij}^1(\Lambda)
\ee
where we have introduced the energy scale $\sigma$ via $ l = { 1 \over \sigma^4}$ and
used the fact that $l=0$ corresponds to the original bare cutoff. We notice
the emergence of the similarity factor $ f_{\sigma ij} = e^{ - {(P^-_i -
P^-_j)^2 \over \sigma^4}} $. 

If we are interested only in particle number conserving (diagonal) part of
the effective interaction, to second order we have,
\be
{d H_{lij} \over dl} = [ \tau_l^{(1)},
H_r^{(1)}]_{ij}
\ee
Using
\be
\tau_{lij}^{(1)} = (P^-_j - P^-_i) H_{rij}^{(1)},
\ee
the effective interaction generated to second order in the diagonal sector is
\be   
H_{lij} = \sum_k H^B_{ik} H^B_{kj} { (P^-_i - P^-_k) + (P^-_j - P^-_k) \over
(P^-_i - P^-_k)^2 + (P^-_j - P^-_k)^2 } \Bigg [ 1 -
e^{ - \Big \{(P^-_i - P^-_k)^2 + (P^-_j - P^-_k)^2 \Big \}/\sigma^4 } \Bigg ].
\ee
Even though the second order formula is very similar to the one in
G{\l}azek-Wilson formalism when an exponential form is chosen for the
similarity factor (see Sec. IV),
we note a slight difference. In the G{\l}azek-Wilson formalism, since the
purpose is to bring the Hamiltonian into a band diagonal form, even in the
particle number conserving sectors the large jumps in energies do not appear
by construction. In the version of the Wegner formalism presented here the
purpose is to bring the Hamiltonian in the block diagonal form in particle
number sector so that large jumps in energies are allowed by the effective
Hamiltonian. Note that small energy denominators do not appear in both
formalisms. 
\section{Effective Hamiltonian in the $q {\bar q}$ sector}
In similarity renormalization approach due to G{\l}azek and Wilson, 
to second order, the interacting part of the
effective Hamiltonian in similarity renormalization
approach is given by Eq. (\ref{gw2eh}). In this work we restrict ourselves
to the $q {\bar q}$ sector. Then the states involved in the matrix elements
$i$ and $j$ refer to $q {\bar q}$ states and $k$ refer to $q{\bar q}g$
states.   

Following the steps similar to the ones outlined in Ref. \cite{dh1}, we 
arrive at the bound state equation 
\be
\Big [ M^2 - {m^2 + k^2 \over x (1-x)} \Big ] \psi_2(x,k) &&= SE \
\psi_2(x,k) \ 
- 4 {g^2 \over 2 (2 \pi)^2}C_f \int dy ~ \int dq ~ f_{\sigma ij}~ \psi_2(y,q) ~{ 1 \over (x-y)^2}
\nonumber \\
&& - {g^2 \over 2 (2 \pi)^2}C_f \int dy ~ \int dq ~\psi_2(y,q)   
{ V \over ED}. \label{ebe1}
\nonumber \\
\ee
Here $x$ and $y$ are the longitudinal momentum fractions and $k$ and $q$ are
the relative transverse momenta. We introduce the cutoff $\eta$ such that 
$ \eta \le x,y \le 1 - \eta$. We further introduce the regulator $\delta$
such that $ \mid x -y \mid \ge \delta$. Ultraviolet divergences are
regulated by the introduction of the cutoff $\Lambda$ on the relative
transverse momenta $k$ and $q$. 
The self energy contribution
\be
SE &&=- {g^2 \over 2 (2 \pi)^2 } C_f \int_0^1 dy \int  dq ~ \theta(x-y)~~
[ 1 - f_{\sigma ik}^2]
~ xy~ { \Big [ \Big ({q \over y} +{ k \over x} - {2 (k-q) \over (x-y)}
\Big )^2 + {m^2 (x-y)^2 \over x^2 y^2} \Big ] \over
(ky-qx)^2 + m^2 (x-y)^2} \nonumber \\
&&~~- {g^2 \over 2 (2 \pi)^2 } C_f \int_0^1 dy \int  dq~\theta(y-x)~
[1 - f_{\sigma ik}^2]
~ (1-x)(1-y) \nonumber \\
&& ~~~~~~~~~~ \times { \Big [ \Big ({q \over 1-y} +{ k \over 1-x} + 
{2 (q-k
) \over (y-x)}
\Big )^2 + {m^2 (y-x)^2 \over (1-x)^2 (1-y)^2} \Big ] \over
[k(1-y)-q(1-x)]^2 + m^2 (x-y)^2} . \label{sei}
\ee 
The boson exchange contribution 
\be
{V \over ED} &&= {\theta (x-y) \over (x-y)} \left [ 
{g_{\sigma jik} \over {m^2 + q^2 \over y}+ {(k-q)^2 \over (x-y)} - {m^2 + k^2 \over x}} 
+ {g_{\sigma ijk} \over {m^2 + k^2 \over 1-x} + {(k-q)^2 \over x-y} - {m^2 +q^2 \over 1-y}}
\right ] \nonumber \\
&& ~~\times \Big [ K(k,x,q,y) ~ + ~i V_I  \Big ]
\nonumber \\
&& + {\theta (y-x) \over (y-x)} \left [ {g_{\sigma jik} \over
{m^2 + k^2 \over x}+ {(q-k)^2 \over (y-x)} - {q^2 +m^2 \over y}}+  
{g_{\sigma ijk} \over {m^2 + q^2 \over 1-y} + {(q-k)^2 \over y-x} - {m^2 +k^2 \over 1-x}}
\right ] \nonumber \\
&& ~~\times \Big [ K(q,y,k,x)~ + ~ i V_I  \Big ] ,
\ee
where
\be
K(k,x,q,y) && =  \left ( {q \over y} + {k \over x} - 
2 {(k-q) \over (x-y)} \right )
\left ( { q \over 1-y} + { k \over 1-x} + {2 (k-q) \over (x-y)} \right )
\nonumber \\
&&~~~~~~~~~- { m^2 (x-y)^2 \over x y (1-x) (1-y)}, \label{bei}
\ee
\be
V_I = - { m \over x y (1-x) (1-y)} [ q (2-y-3x) + k(3y+x-2)].
\ee 
For all the $f$ and $g$ factors, 
\be
M^2_i = {k^2 + m^2 \over x(1-x)} ~ {\rm and} ~
M^2_j = {q^2 + m^2 \over y (1-y)}.
\ee
\be
{\rm For} ~ x > y,~ M_k^2 = {(k-q)^2 \over
x-y} + {q^2 + m^2 \over y} + {k^2 + m^2 \over 1- x} 
\ee
and 
\be {\rm for}~  y > x,
~ M_k^2 = {(q-k)^2 \over y-x} + {q^2 + m^2 \over 1-y} + {k^2 + m^2 \over
x}. 
\ee
 
Before proceeding further, we perform the ultraviolet renormalization. The
only ultraviolet divergent term arises from the factor 1 inside the square
bracket in Eq. (\ref{sei}). We isolate the ultraviolet divergent term
which is given by
\be
SE_{divergent} &&= - {g^2 \over 2 (2 \pi)^2} ~ C_f~ \Bigg [ \int_0^{x -
\delta} dy~
\int_{- \Lambda}^{+ \Lambda} dq ~ {(x+y)^2 \over x y (x-y)^2} \nonumber \\
&& ~~~~~~~~+~ \int_{x + \delta}^1 dy~
\int_{- \Lambda}^{+ \Lambda} dq ~ {(2-x-y)^2 \over (1-x)(1-y) (x-y)^2}
\Bigg ].
\ee
which is canceled by adding a counterterm.   
\section{Similarity Factors}

\subsection{Parameterization I}
In recent numerical work\cite{Allen:2000kx}, the following form for the 
similarity factor has been chosen:
\be
f_{\sigma ij} = e^{ - {{(\Delta M^2_{ij})^2} \over \sigma^4}}
\ee
with $ \Delta {M^2_{ij}} = M^2_i - M^2_j$ where $M^2_i$ denotes the invariant
mass of the state $i$, {\it i.e.}, $M^2_i = \Sigma_i{(\kappa_i^\perp)^2 + m_i^2
\over x_i}$. 
Then 
\be
g_{\sigma ijk} && = f_{\sigma ij} \int_\sigma^\infty d\sigma' ~f_{\sigma' ik}
~ { d \over d \sigma'} f_{\sigma' jk} \nonumber \\
&& = e^{ - {(\Delta M^2_{ij})^2 \over \sigma^4}} ~{(\Delta M^2_{jk})^2 
\over (\Delta
M^2_{ik})^2 + (\Delta M^2_{jk})^2}~ \left [ 1 - 
e^{ - {\Big((\Delta M_{ik}^2)^2 + (\Delta M_{jk}^2)^2\Big)\over \sigma^4}} \right ]~. 
\ee

For the self energy contribution, $i=j$ and we get
\be
g_{\sigma ijk} = g_{\sigma jik} = g_{\sigma iik} ={ 1 \over 2} \left [ 
~1 - e^{ - {2{(\Delta M^2_{ik})^2} \over \sigma^4}} ~ \right ].
\ee
Due to the sharp fall of $f$ with $ \sigma$, the effective Hamiltonian has a
strong dependence on $\sigma$.  Note that this parameterization emerges
naturally in the Wegner formalism.
\subsection{Parameterization II}
St. G{\l}azek has proposed the following form\cite{Glazek:1998sd} 
for $ f_{\sigma ij}$. 
\be
f_{\sigma ij} = { 1 \over \left [1 + \left ({u_{\sigma ij} (1 - u_0) \over
u_0 ( 1 - u_{\sigma ij})} \right )^{2^{n_g}} \right ]} 
\ee
with
\be
u_{\sigma ij} = 
{ {\Delta M^2_{ij}} \over {\Sigma M^2_{ij}} + \sigma^2}, \label{usg}
\ee
$u_0$ a small parameter, and $n_g$ an integer. The {\em mass sum} $ \Sigma
M^2_{ij} = M^2_i + M^2_j$. 
The derivative
\be
{df_{\sigma ij} \over d \sigma} = 2^{n_g} { 2 \sigma \over {\Sigma M^2_{ij}} +
\sigma^2} \left({ u_{\sigma ij} \over u_0}\right)^{2^{n_g}}
{(1 - u_0)^{2^{n_g}} \over ( 1 - u_{\sigma ij})^{2^{n_g}+1}}
{ 1 \over \left [1 + \left ({u_{\sigma ij} (1 - u_0) \over
u_0 ( 1 - u_{\sigma ij})} \right )^{2^{n_g}} \right ]^2}~.
\ee

Note that for small $u$, both $ 1 - f(u)$ and ${d f \over d\sigma}$
vanish like $ u^{2^{n_g}}$.
\subsection{Parameterization III}

For analytical calculations it is convenient to choose\cite{Perry:1994kp} 
a step
function cutoff for the similarity factor:
\be
f_{\sigma ij} = \theta(\sigma^2 - {\Delta M^2_{ij}}).
\ee
Then 
\be
g_{\sigma ijk} = \theta(\sigma^2 - {\Delta M^2_{ij}})~ \theta(
{\Delta M^2_{jk}} - \sigma^2)~ \theta({\Delta M^2_{jk}} - {\Delta M^2_{ik}}).
\ee
It is the factor $ \theta({\Delta M^2_{jk}} - \sigma^2)$ in $g_{\sigma ijk}$
that prevents the energy denominator from becoming small.
\section{Analytical calculations with the step function similarity factor}
In this section we perform analytical calculations to understand the nature
of the effective interactions generated by the similarity factor. Since
there are no divergences associated with $\eta$ and $\Lambda$, we suppress
their presence in the limits of integration in the following equations.

\subsection{Self energy contributions}
Consider the self energy contributions to the bound state equation
Eq. (\ref{sei}). Rewriting the energy denominators to expose the most singular
terms, we have,
\be
SE &&=- {g^2 \over 2 (2 \pi)^2 } C_f \int_0^1 dy \int  dq ~
{\theta(x-\delta -y)
\over x-y}~~
\{ 1 - f_{\sigma ik}^2 \}
 { \Big [ \Big ({q \over y} +{ k \over x} - {2 (k-q) \over (x-y)}
\Big )^2 + {m^2 (x-y)^2 \over x^2 y^2} \Big ] \over
{(k-q)^2 \over (x-y)} +{q^2 +m ^2 \over y} -{k^2 +m^2 \over x}
} \nonumber \\
&&~~- {g^2 \over 2 (2 \pi)^2 } C_f \int_0^1 dy \int  dq~{\theta(y-x-\delta)
\over y-x}~
\{1 - f_{\sigma ik}^2\}
~  \nonumber \\
&& ~~~~~~~~~~ \times { \Big [ \Big ({q \over 1-y} +{ k \over 1-x} + 
{2 (q-k
) \over (y-x)}
\Big )^2 + {m^2 (y-x)^2 \over (1-x)^2 (1-y)^2} \Big ] \over
{(q-k)^2 \over y-x} +{q^2 + m^2 \over 1-y} - {k^2 +m^2 \over 1-x} 
} .
\ee 
The terms associated with 1 in the curly brackets lead to ultraviolet
linear divergent terms which we cancel by counterterms (see Sec. III).  
They also lead to an infrared
divergent term \cite{dh1} which remains uncanceled. 
Explicitly this contribution is given by
\be
 4 {g^2 m^2 \over 2 (2 \pi)^2} ~ C_f ~ \int_0^{x- \delta} dy ~ \int dq
{ 1 \over [ky-qx]^2 + m^2 (x-y)^2} \qquad \qquad \nonumber \\
+ 4 {g^2 m^2 \over 2 (2 \pi)^2} ~ C_f ~ \int_{x+ \delta}^1  dy ~ \int dq
{ 1 \over [k(1-y)-q(1-x)]^2 + m^2 (x-y)^2}. \label{selfir}
\ee 
This is simply
indicative of the fact that terms associated with 1 in the curly bracket
still has a vanishing energy denominator problem. We will address the
resolution of this problem shortly.

Let us next consider new infrared divergences that arise as a result of  
the modifications due to similarity factor.
\subsubsection{Leading singular terms}

Keeping only the most infrared singular terms in the numerators (i.e., for $
x > y$, $4 {(k-q)^2 \over (x-y)^2}$ and for $ y > x$,
$4 {(q-k)^2 \over (y-x)^2}$) and 
denominators (i.e., for $
x > y$, $ {(k-q)^2 \over (x-y)}$ and for $ y > x$,
$ {(q-k)^2 \over (y-x)}$), we have,
\be
SE_1 && =  {g^2 \over 2 (2 \pi)^2 } C_f \int_0^1 dy \int  dq ~ {\theta(x -
\delta -y)
}~  f_{\sigma ik}^2~
4 {1 \over (x-y)^2}  \nonumber \\
&& ~~~    
~~+ {g^2 \over 2 (2 \pi)^2 } C_f \int_0^1 dy \int  dq~\theta(y-x-\delta)~
f_{\sigma ik}^2~ 4 {1 \over (y-x)^2} .
\ee
The
integral is given by 
\be
 \Bigg  [ \int_0^{x - \delta} 
dy \int dq { 1 \over (x-y)^2
} \theta \left(\sigma^2 - {(k-q)^2 \over x-y} \right) + 
\int_{x+ \delta}^1 dy \int dq { 1 \over (y-x)^2
} \theta\left (\sigma^2 - {(k-q)^2 \over y-x} \right) \Bigg  ] .
\ee
We change the transverse momentum variable, $ p=k-q$. For $ x - \delta >y
$, 
we set $ x-y=z$ and for $ y > x + \delta$ we set $ y-x=z$. Then, we have,
\be
4 {g^2 \over 2 (2 \pi)^2 } C_f \left ( \int_\delta^x {dz \over z^2}
\int dp
~ \theta(\sigma^2 - {p^2 \over z}) + 
\int_\delta^{1-x} {dz \over z^2} \int dp 
~\theta(\sigma^2 - {p^2 \over z}) \right ) && = \nonumber \\ 
&& \!\!\!\!\!\!\!\!\!\!\!\!\!\!\!\!\!\!\!\!\!\!\!\!\!\!\!\!\!\!\!\! 
\!\!\!\!\!\!\!\!\!\!\!\!\!\!\!\!\!\!\!\!\!\!
\!\!\!\!\!\!\!\!\!\!\!\!\!\!\!\!\!\!
{16 g^2 \over 2 (2 \pi)^2}~ C_f ~\sigma ~\left [ 
{ 2 \over \sqrt{\delta}} - {1 \over \sqrt{x}} - { 1 \over \sqrt{1-x}}
\right ]. \label{sss} 
\ee
\subsubsection{Sub-leading singular terms}
Next we study sub-leading singular terms containing ${1 \over x-y}$ 
in self energy generated by the
similarity transformation. They are given by 
\begin{eqnarray}
SE_2 &&= - 4 {g^2 \over 2 (2 \pi)^2 } ~C_f \Bigg [ ~\int_0^{x - \delta} 
dy ~ \int dq 
~\theta\left (\sigma^2 - {(k-q)^2 \over x-y} \right ) ~{ 1 \over x-y}~ \Big ( {k^2 \over x} - {q^2 \over
y} \Big ) ~  {1 \over (k-q)^2} \nonumber \\
&& ~~~~~~-  ~ \int_{x+ \delta}^1 dy ~ \int dq 
~ \theta\left (\sigma^2 - {(q-k)^2 \over y-x} \right ) ~{ 1 \over y-x}~ \Big ( {k^2 \over 1-x} - {q^2 \over
1-y} \Big ) ~  {1 \over (q-k)^2} \Bigg ] 
\end{eqnarray}
where we have kept only $(k-q)^2$ term in the denominator since the rest
vanish in the limit $ x \rightarrow y$.
As before, for $ x - \delta > y $, we put $ x-y=z$, $ k-q=p$. With the
symmetric integration in $p$, terms linear in $p$ does not contribute. Only
potential source of $\delta$ divergence is the $p^2$ term in the integrand.
Since $p_{max}= \sigma \sqrt{z}$, after $p$ integration ${ 1 \over z}$ is
converted into ${ 1 \over \sqrt{z}}$ which is an integrable singularity. Same
situation occurs for $ y > x$. Thus 
there are no terms divergent in  $\delta$ coming from sub-leading singular
terms.      
\subsection{Gluon exchange contributions}
Let us next consider the effect of similarity factors on
gluon exchange terms. 
\subsubsection{Instantaneous gluon exchange}
From instantaneous interaction we have,
\be
V_{inst}~= ~- 4 ~{g^2 \over 2 (2 \pi)^2}~ C_f~ \int dy~ \int dq ~ 
\psi_2(y,q)~f_{\sigma ij}~ { 1 \over (x-y)^2}. \label{iinit}
\ee
For the sake of clarity, it is convenient to rewrite this as
\begin{eqnarray}
V_{inst}~ &&=~ -4~ { g^2 \over 2 (2 \pi)^2 }~ {1 \over 2} ~C_f ~\int dy ~
\int dq ~f_{\sigma ij} ~ \psi_2(y,q) \nonumber \\
&& ~~~~\Bigg [ {\theta(x-y - \delta) \over x-y} 
\Bigg \{
{  {(k-q)^2 \over x-y} +({q^2 \over y} -
{k^2 \over x})+ m^2({1 \over y} - { 1 \over x}) 
\over   
{(k-q)^2} +({q^2 \over y} -
{k^2 \over x})(x-y)+ m^2({1 \over y} - { 1 \over x}) (x-y) 
} \nonumber \\
&&~~~~+~{  {(k-q)^2 \over x-y} -({q^2 \over 1-y} -
{k^2 \over 1-x})- m^2({1 \over 1-y} - { 1 \over 1-x}) 
\over   
{(k-q)^2} -({q^2 \over 1-y} -
{k^2 \over 1-x})(x-y)- m^2({1 \over 1-y} - { 1 \over 1-x}) (x-y) 
}
\Bigg \} \nonumber \\
&& ~~~~+ ~ {\theta(y-x - \delta) \over y-x} 
\Bigg \{
{  {(q-k)^2 \over y-x} -({q^2 \over y} -
{k^2 \over x})+ m^2({1 \over x} - { 1 \over y}) 
\over   
{(q-k)^2} -({q^2 \over y} -
{k^2 \over x})(y-x)+ m^2({1 \over x} - { 1 \over y}) (y-x) 
} \nonumber \\
&& ~~~~ + ~ {  {(q-k)^2 \over y-x} +({q^2 \over 1-y} -
{k^2 \over 1-x})+ m^2({1 \over 1-y} - { 1 \over 1-x}) 
\over   
{(q-k)^2} +({q^2 \over 1-y} -
{k^2 \over 1-x})(y-x)+ m^2({1 \over 1-y} - { 1 \over 1-x}) (y-x) 
}
\Bigg \}
\Bigg ].
\ee  
We have to seperately analyze the three types of terms in the numerator.

First consider terms proportional to $m^2$ in the numerator. They are given
by
\be
- 4 ~ {g^2 m^2 \over 2 (2 \pi)^2 } ~ { 1 \over 2} ~C_f ~ \int dy ~\int dq~
f_{\sigma ij} ~\psi_2(y,q) \qquad \qquad \qquad \qquad \qquad \qquad 
\nonumber \\
\times ~\Big [ 
{ 1 \over [ky-qx]^2 + m^2(x-y)^2} + { 1 \over[k(1-y) - q(1-x)]^2 + m^2
(x-y)^2} 
\Big ] \label{coul}
\ee
which leads to the logarithmic confining interaction in the nonrelativistic
limit. Note, however, that Eq. (\ref{coul}) is affected by a logarithmic
infrared divergence arising from the vanishing energy denominator problem.
The logarithmic infrared divergence is cancelled by the self energy
contribution, Eq. (\ref{selfir}). Thus we have explicitly shown that the
logarithmically confining Coulomb interaction survives similarity
transformation but the associated infrared divergence is canceled by self
energy contribution.

Next we look at the most singular term in the numerator in the limit $ x
\rightarrow y$. In this limit we keep only the leading term in the
denominator and we get
\be
- 4 {g^2 \over 2 (2 \pi)^2} ~ C_f~ \int dy ~ \int dq~ \psi_2(y,q) ~
f_{\sigma ij} \left \{ {\theta (x-y) \over (x-y)^2 } + {\theta (y-x) \over
(y-x)^2} \right \}. \label{mostsi}
\ee
Lastly we look at the rest of the terms in the instantaneous exchange. Since
we are interested only in the singularity structure, we keep only the
leading term in the denominator and we get
\be
- 2 {g^2 \over 2 (2 \pi)^2} ~ C_f ~ \int dy ~ \int dq ~ f_{\sigma ij} ~ { 1
\over (k-q)^2} \qquad \qquad \qquad \qquad \qquad \qquad  
\nonumber \\
\times ~ \Bigg [ {\theta (x-y) \over x -y} \Big [{q^2(1-2y) \over y (1-y)} - {k^2(1-2x)
\over x (1-x)} \Big ] 
+ {\theta (y-x) \over y -x} \Big [ {k^2(1-2x) \over x (1-x)} - {q^2(1-2y)
\over y (1-y)} \Big ]. \label{nextsi}
\ee

\subsubsection{Transverse gluon exchange}
First, consider the most singular terms.

Keeping only the most singular terms, the gluon exchange contribution is
\be
&& -{ g^2 \over 2 (2 \pi)^2} ~C_f~ \int dy ~\int dq ~\psi_2(y,q)~ \times
 ~~~~~~~~~~~~
 \nonumber \\
&& \Bigg \{
 {\theta(x- \delta -y) \over x-y} \Bigg [ { g_{\sigma jik} +
g_{\sigma ijk} \over 
{(k-q)^2 \over (x-y)}} (-4) {(k-q)^2 \over (x-y)^2} \Bigg ] 
+ {\theta(y- x - \delta) \over y-x} \Bigg [ { g_{\sigma jik} +
g_{\sigma ijk} \over 
{(q-k)^2 \over (y-x)}} (-4) {(q-k)^2 \over (y-x)^2} \Bigg ] 
\Bigg \}. \nonumber  \\      
&& =  
- { g^2 \over 2 (2 \pi)^2} ~C_f~ \int dy ~\int dq ~\psi_2(y,q) ~ \times 
\qquad \qquad \nonumber \\ 
&& \Bigg \{
 {\theta(x- \delta -y) \over (x-y)^2} \Bigg [  g_{\sigma jik} +
g_{\sigma ijk}  \Bigg ] 
+ {\theta(y- x - \delta) \over (y-x)^2} \Bigg [  g_{\sigma jik} +
g_{\sigma ijk}  \Bigg ] 
\Bigg \} . \nonumber \\
\ee      
Explicitly, for $ x > y$,
\be
g_{\sigma ijk} && =\theta(\sigma^2 - M^2_{ij})~ \theta(M^2_{jk} - M^2_{ik}) ~
\theta (M^2_{jk} - \sigma^2), \nonumber \\
g_{\sigma jik} && =\theta(\sigma^2 - M^2_{ij})~ \theta(M^2_{ik} - M^2_{jk}) ~
\theta (M^2_{ik} - \sigma^2).
\ee
We are interested in the situation $x$ near $y$ and $ i $ near $j$. Then
$ \theta(M^2_{jk} - M^2_{ik}) ={ 1 \over 2} = \theta(M^2_{ik} - M^2_{jk})$
and $ \theta(\sigma^2 - M^2_{ij}) =1$.
Then the gluon exchange contribution is
\be
&& 4 { g^2 \over 2 (2 \pi)^2} ~C_f~ \int dy~ \int dq ~\psi_2(y,q) ~ \times
 \nonumber \\
&& \Bigg [ {\theta(x- \delta -y) \over (x-y)^2} \Bigg \{
1 - \theta \left (\sigma^2 - {(k-q)^2 \over x-y}\right ) \Bigg \}
+ {\theta(y-x- \delta ) \over (y-x)^2} \Bigg \{
1 - \theta \left (\sigma^2 - {(q-k)^2 \over y-x} \right ) \Bigg \} \Bigg ]
\ee   
where we have used $ \theta(x) = 1 - \theta(-x) $.
Combining with the most singular part of the instantaneous contribution
given in Eq. (\ref{mostsi}) we arrive at
\be
&&~~~~~~ - 4 {g^2 \over 2 (2 \pi)^2}~ C_f ~\int dy~ \int dq ~\psi_2(y,q)
~ \times \nonumber \\
&&\Bigg [ {\theta(x- \delta -y) \over (x-y)^2} \theta \left 
( \sigma^2 - {(k-q)^2
\over x-y} \right ) + {\theta(y- x - \delta) \over (y-x)^2} 
\theta \left (\sigma^2 -
{(q-k)^2 \over y-x} \right ) \Bigg ].
\ee
For convenience we change variables. For $ x > y$, we put $ x-y = {p^+ \over
P^+}$ and $k-q = p^1$ and  for $y > x$, we put $y-x = {p^+ \over P^+}$ and 
$ q-k =
p^1$ where $P^+$ is the total longitudinal momentum. Thus we arrive at
\be
&& - 4 {g^2 \over 2 (2 \pi)^2} ~C_f ~P^+~\Bigg [ 
\int dp \int_{P^+\delta}^{P^+x}dp^+ \psi_2(x-{p^+ \over P^+}, k-p^1) 
{ 1 \over (p^+)^2} \theta \left ( \sigma^2  - {(p^1)^2 P^+ \over p^+}
\right ) \nonumber \\
&& ~~~~~~~~~ +  
\int dp \int_{P^+\delta}^{P^+(1-x)}dp^+ \psi_2(x+{p^+ \over P^+}, k+p^1) 
{ 1 \over (p^+)^2} \theta \left (\sigma^2  - {(p^1)^2 P^+ \over p^+} \right ) \Bigg ].
\ee
Consider the Fourier transform
\be
V(x^- , x^\perp) && = - 4 {g^2 \over 2 (2 \pi)^2 }~ C_f~ P^+ 
\Bigg [ 
\int_{P^+ \delta}^{P^{+}x}{dp^{+} \over (p^{+})^2} \int_{-
{p^1}_{max}}^{+p^1_{max}} ~dp^1 ~e^{{i \over 2} p^+ x^- - i p^1 x^1}
\theta \left ( \sigma^2 - {(p^{1})^2 P^{+} \over p^+} \right ) \nonumber \\
&&~~~~~~~ \int_{P^{+} \delta}^{P^+(1-x)}{dp^+ \over (p^+)^2} \int_{-
{p^1}_{max}}^{+{p^1}_{max}} ~dp^1 ~e^{{i \over 2} p^+ x^- - i p^1 x^1}
\theta \left (\sigma^2 - {(p^1)^2 P^+ \over p^{+}} \right ). \Bigg ]
\ee
where $ p^1_{max} = \sigma \sqrt{p^+ \over P^+}$.
We are interested in the behavior of $V(x^-, x^1)$ for  large $x^-, x^1$.
For large $x^-$, nonnegligible contribution to the integral comes from the
region $ q^+ < { 1 \over \mid x^- \mid}$. For large $x^1$, we need $p^1_{max} x^1$ to
be small, i.e., $ (p^1_{max})^2 < { 1 \over (x^1)^2}$, i.e., $p^+ < 
{P^+ \over (x^1)^2 \sigma^2}$. Thus we have the requirements,
$ p^+ < { 1 \over \mid x^- \mid}$, $p^+ < 
{P^+ \over (x^1)^2 \sigma^2}$. We make the approximations 
\be
\int_{-p^1_{max}}^{+p^1_{max}} dp^1 e^{- i p^1 x^1} \approx 2 p^1_{max}
\ee
and $e^{ {i \over 2} q^+ x^- } \approx 1 $.

For large $x^-$, we have $p^+ < { 1 \over \mid x^- \mid} < {P^+ \over (x^1)^2
\sigma^2}$, the upper limit of  $p^+$ integral is cut off by ${ 1 \over
\mid x^- \mid}$. Adding the contributions from both the integrals (which are equal),
for large $x^-$, we have
\be
V(x^-, x^1)~ \approx~ 32~ {g^2 \over 2 (2 \pi)^2} ~ C_f~ \sigma ~\Big [ 
\sqrt{P^+ \mid x^- \mid } - {1 \over \sqrt{\delta}} \Big ]. \label{xlp}
\ee
Thus for large $x^-$ the similarity factors have produced a square root
potential but it is also infrared singular.

For large  $x^1$  the upper limit of $p^+$ integral is cut off by ${P^+
\over (x^1)^2 \sigma^2}$ and we get,
\be
V(x^-, x^1) ~ \approx ~32~ {g^2 \over 2 (2 \pi)^2} ~ C_f~ \sigma ~\Big [ 
\mid x^1 \mid \sigma  - { 1 \over \sqrt{\delta}} \Big ]. \label{xtp}
\ee
For large $x^1$, similarity factors have produced a linear confining
potential which is also infrared singular.
We note that the rotational symmetry is violated in the finite part of the
potential. In both cases, however, the infrared singular part is $
-32 {g^2 \over 2 (2 \pi)^2} ~C_f ~\sigma { 1 \over \sqrt{\delta}}$ which is
exactly canceled by the infrared contribution generated by similarity
transformation from self energy, Eq.
(\ref{sss}).     

Lastly, we consider the terms that go like ${ 1 \over x-y}$. Keeping only
the leading term in the energy denominator, we have,
\be
-2 {g^2 \over 2(2 \pi)^2} ~C_f~ \int dy ~ \int dq ~ \psi_2(y,q)~
\times \qquad \qquad \qquad \qquad \qquad \qquad  \nonumber \\
\Bigg [
{\theta(x-y) \over x-y}~ {g_{\sigma jik} + g_{\sigma ijk} \over (k-q)^2}
\Big [ {k^2(1-2x) \over x (1-x)} - {q^2 (1-2y) \over y (1-y)} \Big ]
\qquad \qquad \qquad \qquad \nonumber \\  
+{\theta(y-x) \over y-x} ~{g_{\sigma jik} + g_{\sigma ijk} \over (q-k)^2}
\Big [ {q^2(1-2y) \over y (1-y)} - {k^2 (1-2x) \over x (1-x)} \Big ]
\Bigg ]. \label {nexttg}
\ee
With the step function cut off we have
\be
g_{\sigma jik} + g_{\sigma ijk} \approx f_{\sigma ij} \theta (\Delta M^2_{ik}
- \sigma^2).
\ee
Then, combining Eq. (\ref{nextsi}) and Eq. (\ref{nexttg}) for the
sub-leading divergences, we get,
\be
 - 2 {g^2 \over 2(2 \pi)^2} ~ C_f~ \int dy ~ \int dq~ ~ f_{\sigma ij}
~ \psi_2(y,q) ~ \times \qquad \qquad \qquad \qquad \qquad \qquad \qquad \qquad \nonumber \\
\Bigg [
{\theta(x-y) \over x-y}~ {1\over (k-q)^2} 
\theta \left (\sigma^2 - {(k-q)^2 \over x-y} \right ) 
\Big [ {k^2(1-2x) \over x (1-x)} - {q^2 (1-2y) \over y (1-y)} \Big ]
\nonumber \\  
+{\theta(y-x) \over y-x} ~{1 \over (q-k)^2}
\theta \left (\sigma^2 - {(q-k)^2 \over y-x} \right ) 
\Big [ {q^2(1-2y) \over y (1-y)} - {k^2 (1-2x) \over x (1-x)} \Big ]
\Bigg ]. 
\ee
Taking the Fourier transform of this interaction, a straightforward
calculation shows that no $log ~\delta$ divergence arise from this
term.  
\subsubsection{Summary of divergence analysis}
The logarithmic confining Coulomb interaction of 2+1 dimensions is 
unaffected by similarity
transformation and is  still affected by a logarithmic divergence which is
however cancelled by a logarithmic divergence from self energy contribution.
Similarity transformation leads to a non-cancellation of the most singular
(${1 \over (x-y)^2}$) term between instantaneous and transverse gluon
interaction terms. This leads to a linear confining interaction for large
transverse seperations and a square root confining interaction for large
longitudinal separations. However, non-cancellation also leads to ${ 1 \over
\sqrt{\delta}}$ divergences where $ \delta $ is the cutoff on $ \mid x-y \mid
$. This divergence is cancelled by new contributions from self energy
generated by similarity transformation. The subleading singular ${ 1 \over x
-y}$ terms do not lead to any divergence in $ \delta$.    
\section{Numerical studies}
The integral equation is converted in to a matrix equation using Gaussian
Quadrature. The matrix is numerically diagonalized using standard LAPACK
routines\cite{laug}. (For details of numerical procedure see Ref. \cite{dh1}.) 
With the exponential form and the step function form of the similarity
factor, the integral over the scale in the definition of $g$ factors in Eq.
(\ref{gfact}) can be performed analytically as shown in Sec. IV. For 
parametrization II, we perform the integration
numerically using $ns$ quadrature points.

The first question we address is the cancellation of divergences which are
of two types: (1) the $ ln ~ \delta$ divergence in the self energy and
Coulomb interaction which has its source in the vanishing energy denominator
problem that survives the similarity transformation and (2) ${ 1
\over \sqrt{\delta}}$ divergences in the self energy and gluon exchange
generated by the similarity transformation. In Table I  we present
the $\delta$ independence of the first five eigenvalues of the Hamiltonian
for $ g$=0.6. Results are presented for three
parametrizations of the similarity factor, namely, the exponential form, the
form proposed by St. G{\l}azek and the step function form used in our
analytical studies. It is clear that the Gaussian Quadrature
effectively achieves the cancellation of $ \delta$ divergences. Recall
that in the study\cite{dh1} of the same problem using Bloch approach,
negative eigenvalues appeared for $g$=.6 when $\delta$ was sufficiently
small (for example, .001) which was caused by the vanishing energy
denominator problem. Our results in the similarity approach for the same
coupling shows that this problem is absent in the latter approach.

Next we study the convergence of eigenvalues with quadrature points. In
Table II we present the results for all three parametrizations of the
similarity factor for $g$=0.2
 with the transverse space discretized using
$k={1 \over \kappa} tan{u \pi \over 2}$ 
where $u$'s are the quadrature
points\cite{dh1}. 
The table show that for $m=1$,
convergence is rather slow for all three choices of the similarity factor
compared to the results in Bloch approach. Among the three choices, 
parameterization II shows better convergence. 

Let us next discuss the nature of low lying levels and wavefunctions. First we 
show the ground state wavefunctions for all three similarity factors for a 
given choice of parameters in Fig. 1. As is anticipated, step function choice
produces a non-smooth wavefunction. For parameterizations I and II,
the wavefunctions show some structure near $x=0.5$. From our previous 
experience with calculations in the Bloch formalism, we believe that the 
structures indicate poor convergence with the number of grid points.

Consider now the structure of low lying levels. Recall that in the 
Bloch formalism, the ordering of levels was $l=0,~1,~0,~ \ldots $ 
corresponding to  logarithmic potential in the nonrelativistic limit. 
In the presence of effective interactions generated by the similarity 
transformation, obviously the level ordering changes. Now we have 
additional confining interactions which, however, act differently in 
longitudinal and transverse directions. The wavefunctions are presented in 
Fig. 2a for parametrization II for the first four low lying levels.

There is extra freedom in parametrization II due to the presence of 
$\Sigma M^2_{ij}$ in the definition of $u_{\sigma ij}$, Eq. (\ref{usg}). For zero 
transverse momentum of constituents, $\Sigma M^2_{ij}$ has the minimum value 
$8m^2$. Thus relative insensitivity of parametrization to $\sigma$ in 
parametrization II for small values of $\sigma$ may be due to this factor. When 
we consider the heavy fermion mass limit, presence of $8m^2$ in 
$u_{\sigma ij}$ enhances the effect of similarity factor. In Fig. 2b we 
present the wavefunctions corresponding to first four levels for 
parametrization II with $8m^2$ subtracted from $\Sigma M^2_{ij}$ in 
$u_{\sigma ij}$. From Figs. 2a and 2b note that the fourth level is different 
for parametrization II with and without $8m^2$ in $u_{\sigma ij}$. 

By suitable choice of parameters we can study the interplay of rotationally 
symmetric logarithmically confining interaction and effective interactions 
generated by similarity transformation. Since for a given $g$, strength of the
logarithmic interaction and similarity generated interactions are 
determined by $m$ and $\sigma$ respectively, for $m>> \sigma$ we should 
recover the Bloch spectrum\cite{dh1}. Upto what levels the recovery occurs, 
of course depends on the exact value of $m$. As we already observed, 
for parametrization II this will happen only if $8m^2$ is subtracted from
$\Sigma M^2_{ij}$. For this case, we present the first four levels for 
$m=10$ and $\sigma =4$ in Fig. 3 which clearly shows the level spacing 
corresponding to the Bloch spectrum.

Finally, we discuss the sensitivity of the spectra to the similarity scale
$\sigma$. Ideally the low lying energy levels should be insensitive to 
$\sigma$. However we have calculated the effective Hamiltonian to only order
$g^2$ and we expect significant sensitivity to $ \sigma$. In Tables III and
IV 
we present
the lowest five eigenvalues for all three parametrizations of the similarity
factor for $g=0.2$ and $g=0.6$ respectively. As expected $\sigma$ dependence
is greater for larger value of $g$. Among the three parametrizations, the 
paramterization II is least sensitive to
$\sigma$. In order to check whether this behaviour is due to the presence of
$8 m^2$ in $\Sigma M^2_{ij}$ in the definition of $u_{\sigma ij}$ we present
the results in Table V for parametrization II with $8 m^2$ subtracted
from $\Sigma M^2_{ij}$. It is clear that sensitivity to $\sigma$ is still
considerably less compared to the other two parametrizations. This may be due
to the fact that  $\Sigma M^2_{ij}$ is added to $\sigma^2$ in the definition
of $\sigma$. Note that in parametrization II sensitivity to $\sigma$ is
controlled also by additional parameters $u_0$ and $n_g$. The sensitivity to
parameters $n_g$ and $u_0$ are presented in Tables VI and VII respectively.

\section{Summary, Discussion and Conclusions}

An attempt to solve 2+1 dimensional gauge theories using the Bloch 
effective Hamiltonian has revealed\cite{dh1} problems due to uncancelled
infrared divergences. They arise out of vanishing energy denominators.
Similarity renormalization formalism attempts to solve the bound state
problem in a two step process. At the first step, high energy degrees of
freedom are integrated out and ultraviolet renormalization carried out
perturbatively. In the second step, the effective Hamiltonian is
diagonalized non-perturbatively. In this work, we have studied the bound
state problem in 2+1 dimensional gauge theories using the similarity
approach.

In order to have a better understanding of the numerical results, we have
performed analytical calculations with step function form for the
similarity factor. Many interesting results emerge from our analytical 
calculations. First of all, it is shown that due to the presence of 
instantaneous interactions in gauge theories on the 
light front, the logarithmic 
infrared divergence that
appeared in the Bloch formalism persists in two places, namely a part of the
self energy contribution and the Coloumb interaction that gives rise to the
logarithmically confining potential in the nonrelativistic limit.
However the terms that persist are precisely those
that produce a cancellation of resulting infrared divergences in the bound
state equation.
The rest of the infrared problem that appeared in the Bloch
formalism due to the vanishing energy denomnator problem is absent in 
the similarity formalism.

Similarity transformation however prevents the cancellation of the most
severe ${1 \over (x-y)^2}$ singularity between instantaneous gluon exchange and
transverse gluon exchange interactions and produces ${ 1 \over
\sqrt{\delta}}$ divergences in the self energy and gluon exchange
contributions which cancells between the two in the bound state equation.
The resulting effective interaction between the quark and antiquark grows
linearly with large transverse separation but grows only with the square
root of the longitudinal separation. This produces violations of rotational
symmetry in the bound state spectrum. We have also verified that no $ln
~ \delta$ divergence result from the ${ 1 \over x-y}$ singularity in the self
energy and gluon exchange contributions.

In the G{\l}azek-Wilson formalism the exact form of the similarity factor
$f_\sigma$ is left unspecified. In the literature an exponential form has
been used in numerical calculations\cite{Allen:2000kx}. For analytical 
calculations it
is convenient to choose a step function even though it is well-known that it
is not suitable for quantitative calculations\cite{Wilson:1974jj}. 
There is also a
proposal due to Stan G{\l}azek which has two extra free parameters.
We have tested all three parametrizations in our work. 
Our numerical results indeed show that step function choice always produces
non-smooth wavefunctions. Parameterization II costs us an extra integration to be
performed numerically but convergence is slightly better for small $g$ compared
to exponential form. All three parametrizations produce violations of
rotational symmetry even for small $g$. When an exponential form is used in
the G{\l}azek-Wilson formalism, the resulting effective Hamiltonian differs
from the Wegner form only by an overall factor that restricts large energy
diffrences between initial and final states. Numerically we have found this
factor to be insignificant.

We have studied the sensitivity of the low lying eigenvalues to the
similarity scale $ \sigma$. Since the effective Hamiltonian is calculated
only to order $g^2$ results do show sensitivity to $\sigma$. Among the three
parametrizations the form II is least sensitive to
$\sigma$ due to the functional form chosen. We have also studied the
sensitivity of eigenvalues to the parameters $u_0$ and $n_g$.  

The bound state equation has three parameters $m$,
$g^2$ and $\sigma$ with dimension of mass. The strength of 
the logarithmically confining interaction
is determined by $m$ and the strength of the rotational symmetry violating
effective interactions generated by similarity transformation is determined
by $\sigma$. For a given $g$ we expect the former to dominate over the 
latter for
$ m >> \sigma$. An examination of low lying eigenvalues and corresponding
wavefunctions show that this is borne out by our numerical calculations.

A major problem in the calculations is the slow convergence.
Compared to the Bloch formalism, in calculations with the
similarity formalism 
various factors may contribute to this problem with the
Gauss quadrature points. One important factor 
is the presence of linear confining interactions generated by the similarity
transformation. It is well known that such interactions are highly singular
in momentum space. Another factor is the presence of ${ 1 \over
\sqrt{\delta}}$ divergences the cancellation of which is achieved
numerically. It is of interest to carry out the same calculations with
numerical procedures other than the Gauss quadrature. However, one should
note that calculations in 3+1 dimensions employing basis functions and
splines have also yielded\cite{Allen:2000kx} wavefunctions which show 
non-smooth structures.

An undesirable result of the similarity transformation carried out in
perturbation theory is the
violation of rotational symmetry. Our results show that this  violation
persists at all values of $g$ for $m=1$. Such a violation was also observed 
in 3+1
dimensions. In that case the functional form of the logarithmic potential
generated by similarity transformation is the same in longitudinal and
transverse directions but the coefficients differ by a factor of two. Same
mechanism in 2+1 dimensions makes even the functional forms different. The
important questions are whether the confining interactions generated by the
similarity transformation are an artifact of the lowest order approximation
and if they are not, then, whether the violation of rotational symmetry will
diminish with higher order corrections to the effective Hamiltonian.  
Recall that
the high energy degrees of freedom has been integrated out and the effective
low energy Hamiltonian determined only to order $g^2$. A clear answer will
emerge only after the determination of the effective Hamiltonian to fourth
order in the coupling.        

\eject
\vskip .5in
\figure{\noindent FIG. 1. The ground state wavefunction  
for different choices of the similarity factor using the tan parametrization
for transverse momentum grid and for $n_1=40$, $n_2=80$, $m$=1.0, $g=0.2$, $\sigma=4.0$ and $
\eta = \delta =10^{-5}$, $\kappa=10.0$ 
as a function of $x$ and $k$. (A) Parametrization I, (B) Parametrization II
with $n_g=2$, $u_0=0.1$, $ns=500$,
(C) Parametrization III. }
\vskip .3in
\figure{\noindent FIG. 2a. The wavefunctions corresponding to the lowest
four
eigenvalues  as a function of $x$ and $k$ for parametrization II.
The parameters are as in FIG. 1. (A) Lowest state, (B) first excited
state, (C) second excited state, (D) third excited state. }
\vskip .3in
\figure{\noindent FIG. 2b. Same as in FIG. 2a but with $8m^2$ subracted from 
$ \Sigma M^2_{ij}$ .}
\vskip .3in
\figure{\noindent FIG. 3. The wavefunctions corresponding to the lowest four
eigenvalues as a function of $x$ and $k$ with parametrization III with
$8m^2$ subtracted from $\Sigma M^2_{ij}$.
The parameters are
$g=0.2$, $\eta=\delta=.00001$,
$m=10.0$, $\kappa=10.0$, $n_g=2$, $u_0=0.1$, $n_1=40$, $n_2=80$.  
(A) Lowest state, (B) first excited
state, (C) second excited state, (D) third excited state. }
\eject
\vskip 1in
\begin{tabular}{||c|c|c|c|c|c||}
\hline \hline
 $\delta$   &  \multicolumn{5}{c||}  {Parametrization I} \\
\hline                                                           
     0.1 & 4.89535 & 4.90359 & 4.90420 & 4.90420 & 4.90482 \\
     0.01 & 5.62612 & 6.38083 & 6.82963 & 7.20037 & 7.36414 \\
     0.001 & 5.68417 & 6.42147 & 6.90879 & 7.30609 & 7.64650 \\ 
     0.0001 & 5.68432 & 6.42148 & 6.90909 & 7.30611 & 7.64677 \\
     0.00001 & 5.68432 & 6.42148 & 6.90909 & 7.30611 & 7.64677 \\
\hline
   $\delta$   &  \multicolumn{5}{c||}  {Parametrization II} \\
\hline
      0.1 & 4.55364 & 4.55668 & 4.55668 & 4.55668 & 4.55669 \\
      0.01 & 4.86066 & 5.33491 & 5.49693 & 5.59838 & 5.79111 \\
      0.001 & 4.87607 & 5.35671 & 5.59226 & 5.64476 & 5.88613 \\
      0.0001 & 4.87604 & 5.35671 & 5.59236 & 5.64477 & 5.88615 \\
     ~ 0.00001~ & ~4.87604~ & ~5.35671~ & ~5.59236~ & ~5.64477~ & 
    ~5.88615~ \\ 
\hline
 $\delta$   &  \multicolumn{5}{c||}  {Parametrization III} \\
\hline
     0.1 & 5.12410 & 5.13039 & 5.13101 & 5.13101 & 5.13754 \\
     0.01 & 6.02600 & 6.94326 & 7.50445 & 7.98054 & 8.39927 \\
     0.001 & 6.00968 & 6.97160 & 7.55376 & 8.07749 & 8.49199 \\
     0.0001 & 5.96636 & 6.97160 & 7.51814 & 8.07751 & 8.46524 \\
     0.00001 & 5.96636 & 6.97160 & 7.51814 & 8.07751 & 8.46524 \\
\hline
\hline
\end{tabular}
\vskip 1cm
{TABLE I. Variation with $\delta$ of the first five eigenvalues of the 
 full hamiltonian (excluding the less
significant imaginary term). The parameters
are $m=1.0$,
$g=0.6$, $n_1=58$, $n_2=58$, $\eta=0.00001$, $ \Lambda=20.0$, $\sigma=4.0$,
($u_0=0.1,$ $n_g=2$ and $ ns=500$ (for $\sigma$ integration) in parametrization
II)}

\vskip 1in
\begin{tabular}{||c|c|c|c|c|c|c||}
\hline \hline
$n_1$ &  $n_2$ &   \multicolumn{5}{c||}{Parametrization I     } \\
\hline
  ~10~ & ~10~ & ~4.320~ & ~4.353~ & ~4.357~ & ~4.357~ & 
  ~4.361~ \\
  20 & 20 & 4.375 & 4.442 & 4.484 & 4.484 & 4.485 \\
  20 & 30 & 4.398 & 4.482 & 4.546 & 4.583 & 4.610 \\
  20 & 40 & 4.411 & 4.502 & 4.570 & 4.615 & 4.656 \\
  30 & 40 & 4.412 & 4.503 & 4.570 & 4.615 & 4.655 \\
  40 & 50 & 4.420 & 4.515 & 4.585 & 4.634 & 4.678 \\
  40 & 60 & 4.426 & 4.524 & 4.594 & 4.645 & 4.692 \\
  40 & 80 & 4.434 & 4.535 & 4.607 & 4.661 & 4.709 \\  
\hline
 $n_1$ &  $n_2$ & 
\multicolumn{5}{c||}   {Parametrization II}\\
\hline 
 ~~10~~ & ~~10~~ & ~4.163~ & ~4.194~ & ~4.194~ & ~4.194~ &
 ~4.203~\\
 ~20~ &  ~20~ & ~4.186~ & ~4.244~ &  ~4.276~ & ~4.276~ &
 ~4.277~ \\
  ~20~ &  ~30~ & 4.192 & 4.256 & 4.296 & 4.323 &
 4.329 \\
  20 &  40 & 4.195 & 4.262 & 4.304 & 4.335 &
 4.344 \\
  30  &  40  & 4.195 & 4.262 & 4.304  & 4.335 &
 4.344 \\
  40 & 50 &  4.197 & 4.266 & 4.308 & 4.341 & 4.353 \\
  40 & 60 & 4.199 & 4.268 & 4.311 & 4.345 & 4.359  \\
  40 & 80 & 4.201 & 4.272 & 4.315 & 4.350 & 4.367 \\
 
\hline
$n_1$ &  $n_2$ &   \multicolumn{5}{c||}{Parametrization III} \\
\hline
 10 & 10 & 4.360 & 4.379 & 4.381 & 4.381 & 4.391 \\
 20 & 20 & 4.469 & 4.533 & 4.572 & 4.572 & 4.572 \\
 20 & 30 & 4.503 & 4.604 & 4.674 & 4.721 &  4.749 \\
 20 & 40 & 4.520 & 4.632 & 4.703 & 4.768 & 4.810 \\
 30 & 40 & 4.528 & 4.636 & 4.714 & 4.768 & 4.811 \\
 40 & 50 & 4.542 & 4.657 & 4.736 & 4.797 & 4.848 \\
 40 & 60 & 4.548 & 4.668 & 4.748 & 4.813 & 4.866 \\
 40 & 80 & 4.556 & 4.683 & 4.764 & 4.833 & 4.889\\
\hline
\hline
\end{tabular}
\vskip 1cm
{TABLE II.
 Convergence of eigenvalues with $n_1$ and $n_2$ for the parametrization
$k={1\over \kappa}tan(q\pi/2)$. The parameters
are 
 $m$=1.0, $g$=0.2, $\eta=0.00001$, $\delta=0.00001$, $\kappa=10.0$,
$\sigma=4.0$, ($u_0$=0.1, $n_g$=2 and the number of quadrature points $ns=500$
for 
$\sigma$ integration for parametrization II).
   }
\vskip 1in
\begin{tabular}{||c|c|c|c|c|c||}
\hline \hline
& \multicolumn{5}{c||}{Eigenvalues ($M^2$)} \\
\hline
 $\sigma$   &  \multicolumn{5}{c||}  {Parametrization I} \\
\hline                                                           
     2.0 & 4.214 & 4.285 & 4.329 & 4.365 & 4.393\\ 
     4.0 & 4.434 & 4.535 & 4.607 & 4.661 & 4.709\\
     6.0 & 4.701 & 4.821 & 4.914 & 4.980 & 5.043\\
\hline
   $\sigma$   &  \multicolumn{5}{c||}  {Parametrization II} \\
\hline
      ~2.0~ & ~4.157~ & ~4.214 ~ & ~4.248~ & ~4.260 ~ & ~4.276~ \\
      4.0 & 4.201 & 4.272 & 4.315 & 4.350 & 4.367 \\
      6.0 &  4.266 & 4.350 & 4.404 & 4.446 & 4.483 \\
\hline
 $\sigma$   &  \multicolumn{5}{c||}  {Parametrization III} \\
\hline
      2.0 & 4.254  & 4.346  & 4.395  & 4.443 & 4.477 \\
      4.0 & 4.556 & 4.683  & 4.764  & 4.833 & 4.889  \\
      6.0 & 4.927 & 5.075 & 5.179  & 5.262 & 5.333 \\

\hline
\hline
\end{tabular}
\vskip 1cm
{TABLE III. Variation with $\sigma$ of the  full hamiltonian (excluding the 
 imaginary term). The parameters
are $m=1.0$,
$g=0.2$, $n1=40$, $n2=80$, $\eta=0.00001$, $ \kappa=10.0$,
$(k={1\over
\kappa}tan(q\pi/2))$, $\delta=0.00001$,
($u_0=0.1$, $n_g=2$, and $ ns=500$ for $\sigma$ integration for
parameterization II)}
\eject
\vskip 1in
\begin{tabular}{||c|c|c|c|c|c||}
\hline \hline
& \multicolumn{5}{c||}{Eigenvalues ($M^2$)} \\
\hline
 $\sigma$   &  \multicolumn{5}{c||}  {Parametrization I} \\
\hline                                                           
     2.0 & 4.941 & 5.418 & 5.708 & 5.713 & 5.950 \\ 
     4.0 & 5.684 & 6.421 & 6.909 & 7.306 & 7.647 \\
     6.0 & 6.678 & 7.617 & 8.274 & 8.801 & 9.267 \\
\hline
   $\sigma$   &  \multicolumn{5}{c||}  {Parametrization II} \\
\hline
     ~ 2.0~ &~ 4.769 ~& ~5.152~ & ~5.226~ &~ 5.372~ &~ 5.485~ \\
      4.0 & 4.876 & 5.357 & 5.592 & 5.645 & 5.886 \\
      6.0 &  5.071 & 5.655 & 6.016 & 6.159 & 6.316 \\
\hline
 $\sigma$   &  \multicolumn{5}{c||}  {Parametrization III} \\
\hline
      2.0 & 4.888  & 5.589  & 5.882  & 6.120 & 6.263 \\
      4.0 & 5.966 & 6.972  & 7.518  & 8.077 & 8.465  \\
      6.0 & 7.359 & 8.603 & 9.360  & 10.083 & 10.621 \\

\hline
\hline
\end{tabular}
\vskip 1cm
{TABLE IV. Variation with $\sigma$ of the  full hamiltonian (excluding the 
 imaginary term). The parameters
are $m=1.0$,
$g=0.6$, $n1=58$, $n2=58$, $\eta=0.00001$, $ \Lambda=20.0$, 
$(k={q\Lambda m
\over (1-q^2)\Lambda +m})$, $\delta=0.00001$,
($u_0=0.1$, $n_g=2$,and $ ns=500$ for $\sigma$ integration for
parameterization II)}
\vskip 1in
\eject
\begin{tabular}{||c|c|c|c|c|c|c||}
\hline \hline
& & \multicolumn{5}{c||}{Eigenvalues ($M^2$)} \\
\hline
  $g$ & $\sigma$   &  \multicolumn{5}{c||} {Parametrization II} \\
\hline
        & 2.0 & 4.126 & 4.166 & 4.176 & 4.189 & 4.201 \\
  ~0.2~ &~4.0~ & ~4.172~ & ~4.235~ &~ 4.273~ & ~4.298~ & ~4.304~ \\
     & 6.0 &  4.241 & 4.321 & 4.371 & 4.411 & 4.445 \\
\hline
\hline
      & 2.0 & 4.739 & 5.003 & 5.020 & 5.134 & 5.193 \\ 
  0.6  & 4.0 & 4.802 & 5.222 & 5.350 & 5.470 & 5.626 \\
    & 6.0 & 4.993 & 5.541 & 5.876 & 5.940 & 6.156 \\
\hline
\hline
\end{tabular}
\vskip 1cm
{TABLE V. Variation with $\sigma$ of the  full hamiltonian (excluding the 
 imaginary term) after subtracting $8m^2$ from $\Sigma M^2_{ij}$ in
the definition of $u_{\sigma ij}$.
 The parameters
are 
 $m=1.0$, $\eta=0.00001$, $\delta=0.00001$,
$u_0=0.1$, $n_g=2$, and $ ns=500$  \\
 1) for $g=0.2$, $k={1\over\kappa}tan(q\pi/2)$ with $ \kappa=10.0$ and 
$n1=40$, $n2=80$\\
 2) for $g= 0.6$, $k={q\Lambda m \over (1-q^2)\Lambda +m}$ with
$\Lambda=20.0$, and $n1=n2=58$. }
\eject
\vskip 0.5cm
\begin{center}
\begin{tabular}{||c|c|c|c|c|c|c||}
\hline \hline
  $g$ & $\sigma$   &  \multicolumn{5}{c||}  {$M^2$
(Parameterization II)} \\
\hline
\hline
      & 2.0 & 4.781 & 5.049 & 5.051 & 5.161 & 5.221 \\ 
  0.6  & 4.0 & 4.843 & 5.235 & 5.391 & 5.464 & 5.639 \\
    & 6.0 & 5.030 & 5.538 & 5.848 & 5.956 & 6.106 \\
\hline
\hline
\end{tabular}
\end{center}
\vskip 1cm
{TABLE VI. Variation with $\sigma$ of the  full hamiltonian(excluding the 
 imaginary term) after subtracting the $8m^2$ term from $\Sigma M^2_{ij}$ in 
the definition of $u_{\sigma ij}$.
 The parameters
are 
 $m=1.0$, $\eta=0.00001$, $\delta=0.00001$,
$u_0=0.1$, $n_g=1$, and $ ns=500$  
  $g= 0.6$, $k={q\Lambda m \over (1-q^2)\Lambda +m}$ with
$\Lambda=20.0$, and $n1=n2=58$. }
\vskip 1cm
\vskip0.5cm
\begin{center}
\begin{tabular}{||c|c|c|c|c|c|c||}
\hline \hline
  $g$ & $u_0$   &  \multicolumn{5}{c||}  {$M^2$  (Parameterization II)} \\
\hline
     & 0.2 & 4.990 & 5.548 & 5.887 & 5.969 & 6.171\\
     & 0.1 & 4.802 & 5.222 & 5.350 & 5.469 & 5.626 \\
 0.6 & 0.05 & 4.746 & 5.076 & 5.091 & 5.253 & 5.313 \\
     & 0.01 & 4.813 & 5.003 & 5.029 & 5.069 & 5.130 \\
\hline
\hline
\end{tabular}
\end{center}
\vskip 1cm
{TABLE VII. Variation with $u_0$ of the  full hamiltonian(excluding the 
 imaginary term) after subtracting the $8m^2$ term from 
$\Sigma M^2_{ij}$ the definition of $u_{\sigma ij}$.
The parameters are $m=1.0$, $\eta=0.00001$, $\delta=0.00001$,
$\sigma=4.0$, $n_g$=2, and $ ns=500$  
  $g= 0.6$, $k={q\Lambda m \over (1-q^2)\Lambda +m}$ with
$\Lambda=20.0$, and $n1=n2=58$. }

\begin{figure}
\begin{center}
\psfig{figure=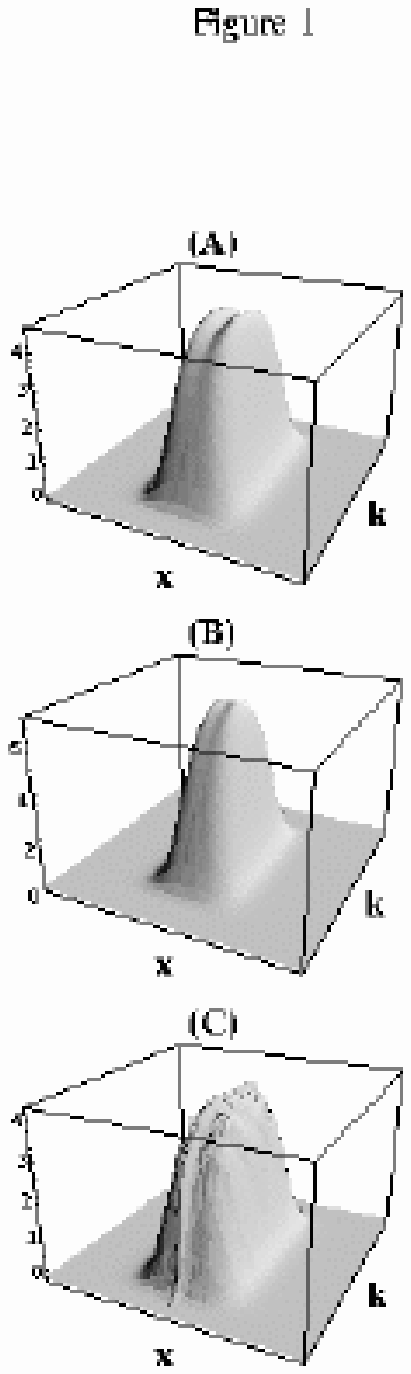,width=16.0cm,height=16.0cm}
\end{center}
\end{figure}
\begin{figure}
\begin{center}
\psfig{figure=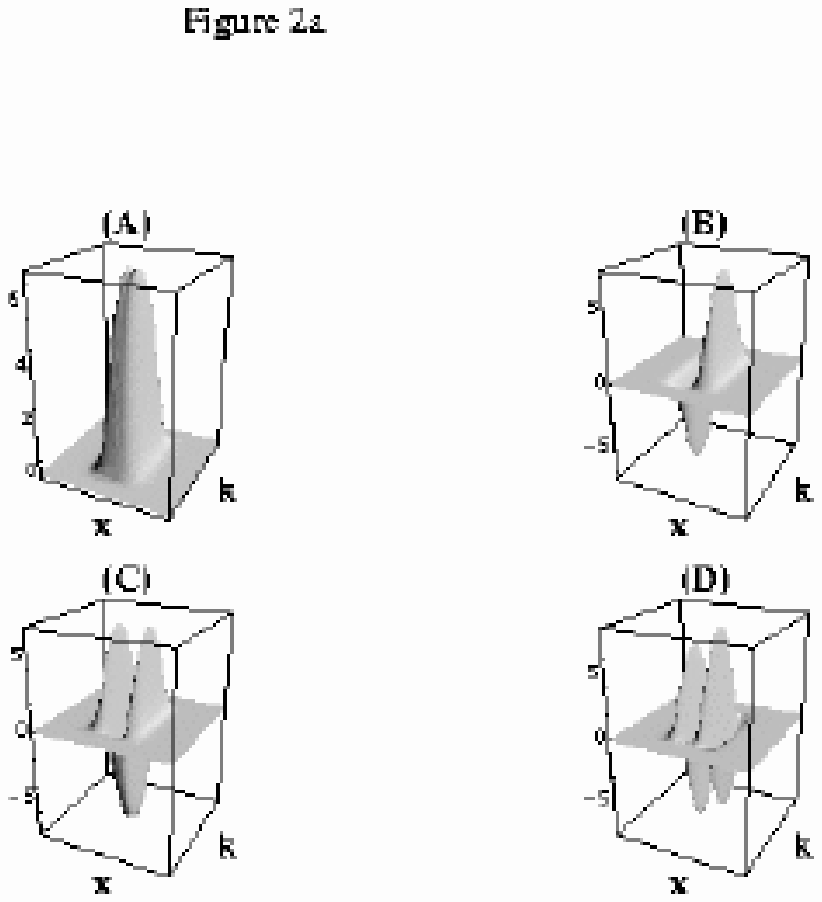,width=16.0cm,height=16.0cm}
\end{center}
\end{figure}
\begin{figure}
\begin{center}
\psfig{figure=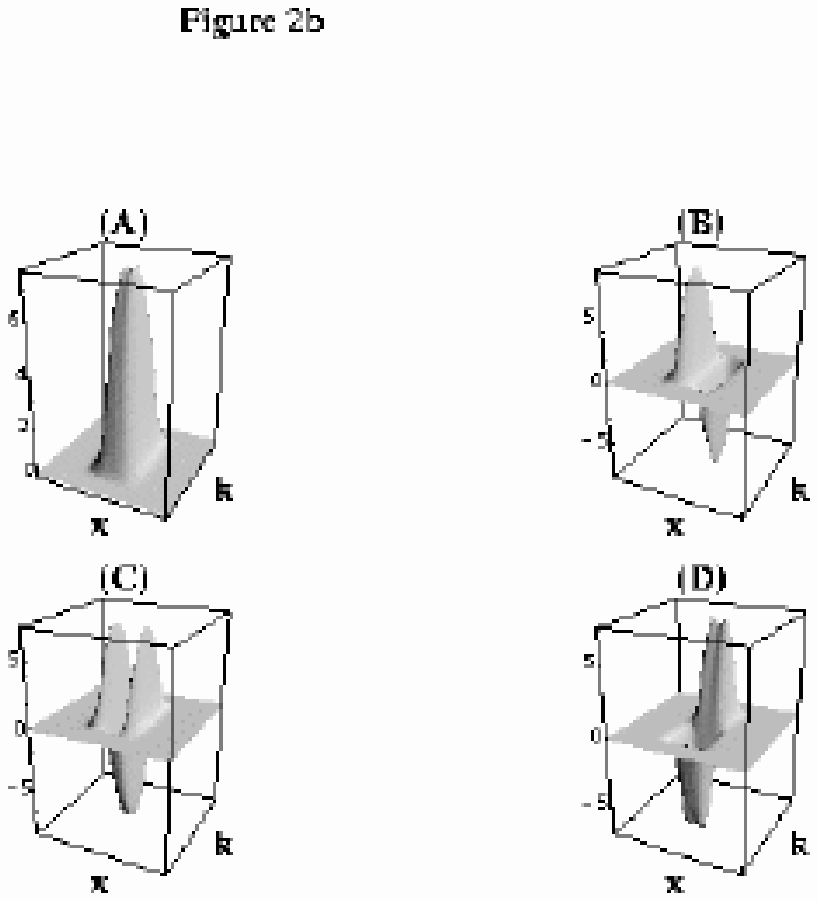,width=16.0cm,height=16.0cm}
\end{center}
\end{figure}
\begin{figure}
\begin{center}
\psfig{figure=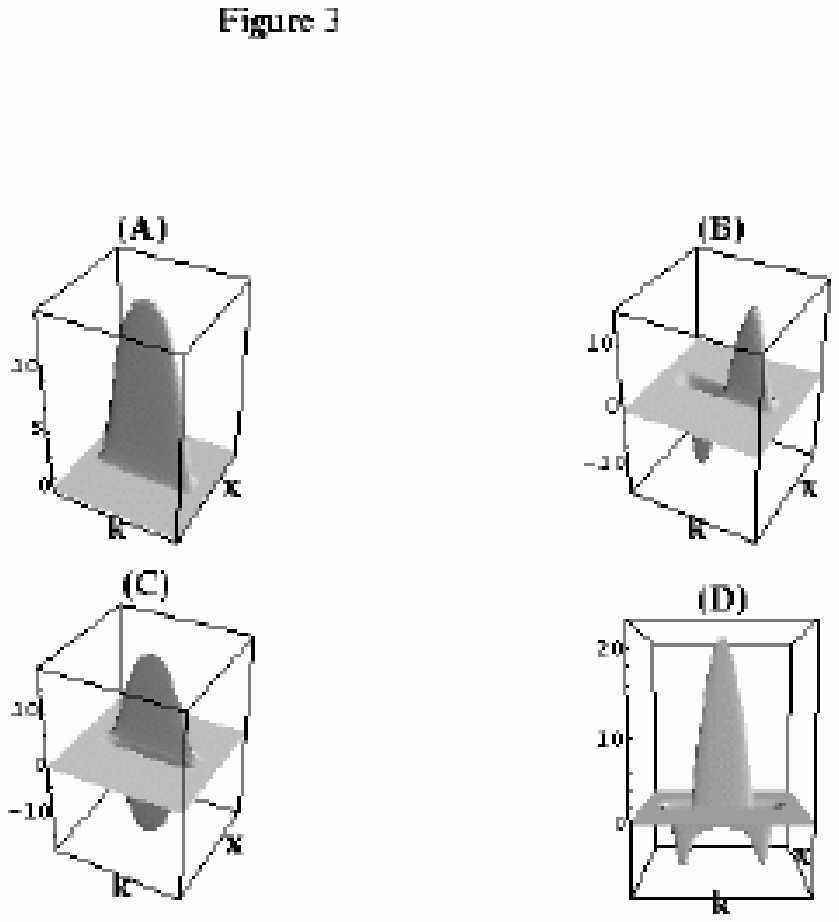,width=16.0cm,height=16.0cm}
\end{center}
\end{figure}
\end{document}